\documentclass{article}
\usepackage{spconf,amsmath,epsfig, hyperref}
\usepackage[utf8]{inputenc}
\usepackage{graphicx}
\usepackage{subfigure}
\usepackage{xcolor}
\usepackage{booktabs}
\usepackage{placeins}
\usepackage{hyperref}
\usepackage{comment}
\usepackage{cite}
\usepackage{balance}

\usepackage{url}

\title{A Hybrid MLP-Quantum Approach in Graph Convolutional Neural Networks for Oceanic Niño Index (ONI) Prediction}

\name{\begin{tabular}{c}Francesco Mauro$^{a,1}$, Alessandro Sebastianelli$^{b}$, Bertrand Le Saux$^{b}$, \\ Paolo Gamba$^{c}$
\textit{and Silvia Liberata Ullo}$^{a}$\end{tabular}\thanks{
$^{1}$Corresponding author. 
\textit{Email addresses}: f.mauro$@$studenti.unisannio.it (FM), alessandro.sebastianelli$@$esa.int (AS), bertrand.le.saux$@$esa.int (BLS), paolo.gamba@unipv.it (PG), ullo$@$unisannio.it (SLU)} 
}

\address{
$^{a}$ Engineering Department, University of Sannio, Benevento, Italy \\
$^{b}$ $\phi$-lab, European Space Agency, Frascati, Italy \\
$^{c}$ Engineering Department, University of Pavia, Pavia, Italy 
}

\begin{document}
\maketitle

\begin{abstract}
This paper explores an innovative fusion of Quantum Computing (QC) and Artificial Intelligence (AI) through the development of a 
Hybrid Quantum Graph Convolutional Neural Network
(HQGCNN), combining a Graph Convolutional Neural Network (GCNN) with a Quantum Multilayer Perceptron (MLP). 
The study highlights the potentialities of GC\nobreak NNs in handling global-scale dependencies and proposes the HQGCNN for predicting complex phenomena such as the Oceanic Niño Index (ONI). 
Preliminary results suggest the model potential to surpass state-of-the-art (SOTA). 
The code will be made available with the paper publication.
\end{abstract}

\section{Introduction}
\label{sec:intro}
In recent years, the intersection of Quantum Computing (QC) and Artificial Intelligence (AI) has yielded innovative approaches to tackle complex computational problems. One promising avenue in this realm is the development of Hybrid Quantum Neural Networks (HQNNs), where classical and quantum components collaborate synergistically to harness the advantages of both paradigms. Within this rapidly evolving landscape, the integration of quantum mechanics into neural network architectures stands out as a captivating frontier.

An HQNN amalgamates classical neural network structures, such as Multilayer Perceptrons (MLPs), with QC elements to exploit the inherent parallelism and computational prowess offered by quantum states. This amalgamation presents a novel paradigm for information processing, promising breakthroughs in solving intricate problems that transcend the capabilities of classical computing alone \cite{ullo2024enhancing}.
\\
\noindent With the advent of Quantum Machine Learning (QML), deep research has been dedicated by the authors to the examination and assessment of suitable quantum algorithms for addressing contemporary challenges in Earth Observation (EO). In \cite{zaidenberg2021advantages}, a proof-of-concept for an HQNN applied to EO land-use-land-cover (LULC) binary classification has been introduced. This initial model employed a simple hybrid architecture with a limited number of qubits and a single circuit layer. Building upon this earlier work, in \cite{sebastianelli2021circuit}, the solution was enhanced to enable multiclass classification. The analysis involved exploring various circuits, conclusively demonstrating the superior performance of real amplitudes through the exploitation of quantum entanglement.
In \cite{sebastianelli2023quantum}, guidelines for tuning hyperparameters within the quantum component have been produced. This effort was geared towards demonstrating the quantum advantage in the realm of RS data classification, and showcasing that there are more advantageous solutions than merely increasing the number of qubits. 
Configurations with increased layers and qubits, as well as alternative layer types, including quantum convolution operations, have been delved.
However, although Quantum Convolutional Neural Networks (QCNNs) excel at capturing spatial features, they exhibit limitations in capturing global-scale dependencies or events, such as El Niño fluctuations.
Motivated by this observation, the research presented in this paper aims to extend the exploration of 
Hybrid Quantum Graph Convolutional Neural
Networks (HQGCNNs), combining a Graph Convolutional
Neural Network (GCNN) with a Quantum Multilayer Per-
ceptron (MLP), by demonstrating the ability to learn the characteristics of spatially distant relationships between EO measurements and  Earth phenomena varying in time.\\
The paper is organized as follows: in the next section, the El Niño phenomenon is described with some of the most common methods applied for its monitoring. Then the dataset used for this study is introduced. After that, the proposed method is presented in detail, and the preliminary obtained results are explained and compared with state-of-the-art (SOTA). The Discussion and Conclusion section ends the paper. 

\section{El Niño fluctuations}
\label{sec:fluctuation}
\begin{figure*}[!ht]
	\centering
    \includegraphics[width=2\columnwidth]{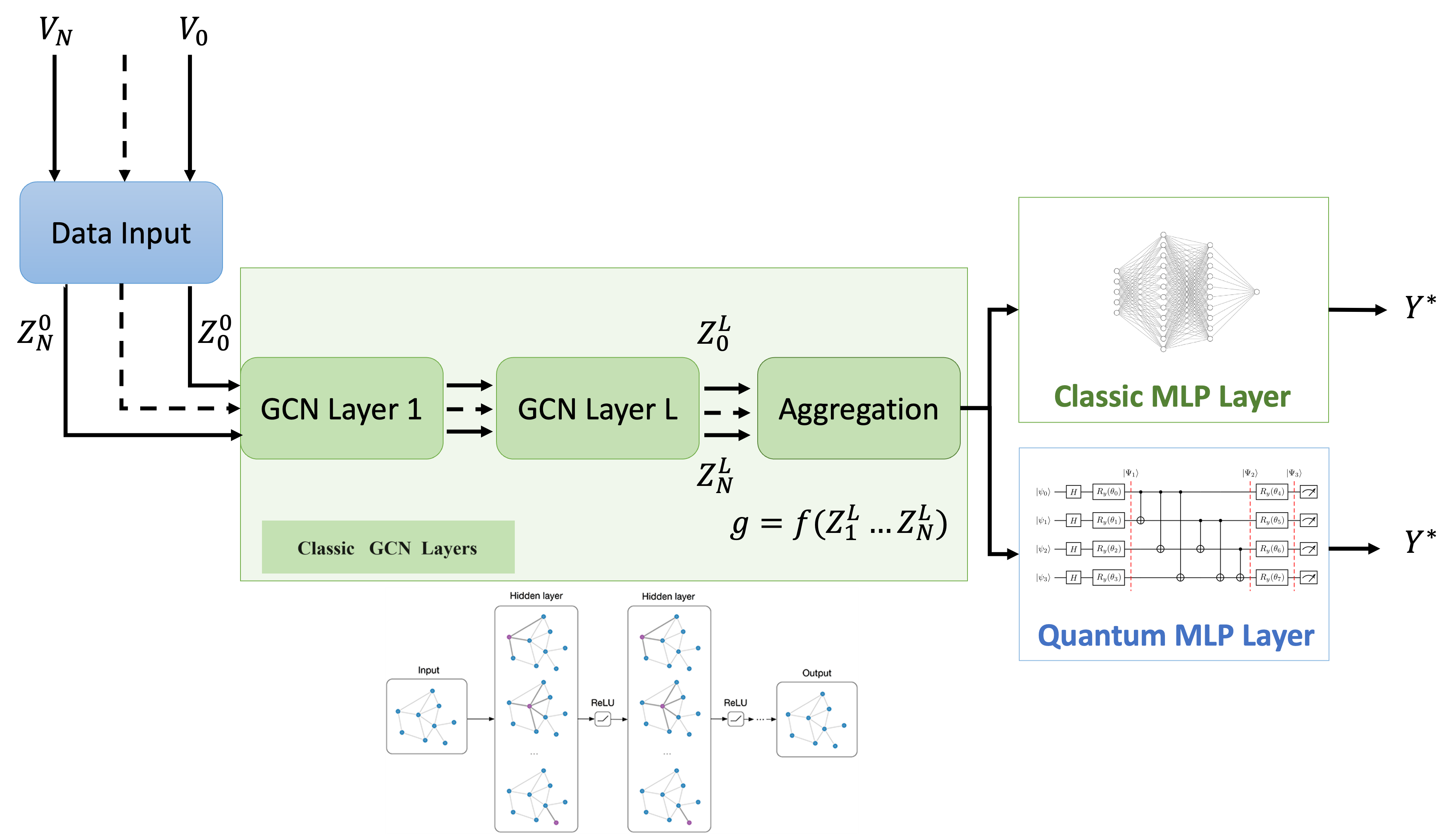}
	\caption{Architecture of the proposed Hybrid Quantum Graph Convolutional Neural Network.}
	\label{QGNN_arch}
\end{figure*}
El Niño fluctuations, also known as El Niño-Southern Oscillation (ENSO), are a recurring climate phenomenon characterized by an abnormal warming of the central and eastern tropical Pacific Ocean. The occurrence and intensity of El Niño fluctuations are monitored and studied by various scientific organizations and research institutions. Many works provide detailed explanations and insights into El Niño phenomenon, the mechanisms of its fluctuations, and their broader impacts \cite{McPhaden2012, Glantz2002, IPCC2013}. 
These works serve as valuable resources for understanding the dynamics of El Niño events and their significance in the Earth's climate system. 
Based on findings from the literature, seasonal and long-range forecast\nobreak ing has appeared to be a suitable application for Graph Neural Networks (GNNs) and GCNNs \cite{hamilton2020graph}, rather than CNNs, due to their translational equivariance caused by parameter-shared convolutions \cite{cachay2021world}, \cite{heaton2018ian}. In the following, the main limitations of CNNs are highlighted to stress the importance of a choice towards the GNNs and GCNNs:

\begin{itemize}
    \item The convolutional layers of CNNs are shift-equivariant:  it means that a shift of the input to a convolutional layer produces a shift in the output feature maps by the same amount \cite{bronstein2021geometric}. However, in EO  applications, the specific location of patterns holds significant importance. 
    \item CNNs create representations by focusing on nearby regions in input, resulting in a bias towards spatial proximity. Nonetheless, numerous climate phenomena are influenced by global interactions. 
    \item CNNs encounter inflexibility when it is not required to use all grid cells of the input, even in cases where certain regions of the input are known to be unnecessary and should be excluded from the CNN's analysis.
\end{itemize}

Concerning the GNNs, while over the past years, their use in earth and atmospheric sciences has been limited \cite{saha2022multitarget}, for some applications their employment has become attractive and successful, such as earthquake source detection \cite{bilal2022early}, power outage prediction \cite{owerko2018predicting}, and wind-farm power estimation \cite{park2019physics}. In addition, the possibility of representing data as a graph makes GNNs highly promising for capturing distant relationships in many phenomena such for instance, the ENSO forecasting  \cite{cachay2021world}. \\
All of the above has pushed us to study and propose an HQGCNN, combining a GCNN with a Quantum MLP. The idea is to merge the advantages of GCNNs and QC. The proposed HQGCNN is 
schematized in Fig. \ref{QGNN_arch}. The description of the architecture is given in detail in the section 4. Methods and Results. 


\section{Dataset}
\label{sec:Dataset}
Based on the work proposed in \cite{cachay2021world}, the datasets we used are: 1) the SODA reanalysis dataset (1871-1973) as a Train set, and 2) the GODAS dataset for the period of 1984 to 2017 as a Test set. In particular, together with the SODA dataset, climate model simulations from Coupled Model Intercomparison Project Phase five (\href{https://pcmdi.llnl.gov/mips/cmip5/}{CMIP5}) have also been used. Through the augmentation of the dataset with potentially noisy simulations,  a sample size suitable for deep learning methods is reached (in total 30k samples). It is worth highlighting that the datasets are used in a resolution of 5 degrees, and only the geographical locations that lie within $55S-60N$ and $0-360W$ are used.
The prediction target is the  Oceanic Niño Index (ONI) index for h months ahead. ONI represents the NOAA's primary index for tracking the ocean part of ENSO, and ENSO climate patterns. 

\section{Method and results}
\label{sec:method}
\begin{figure}[!ht]
	\centering    
    \includegraphics[scale=0.42]{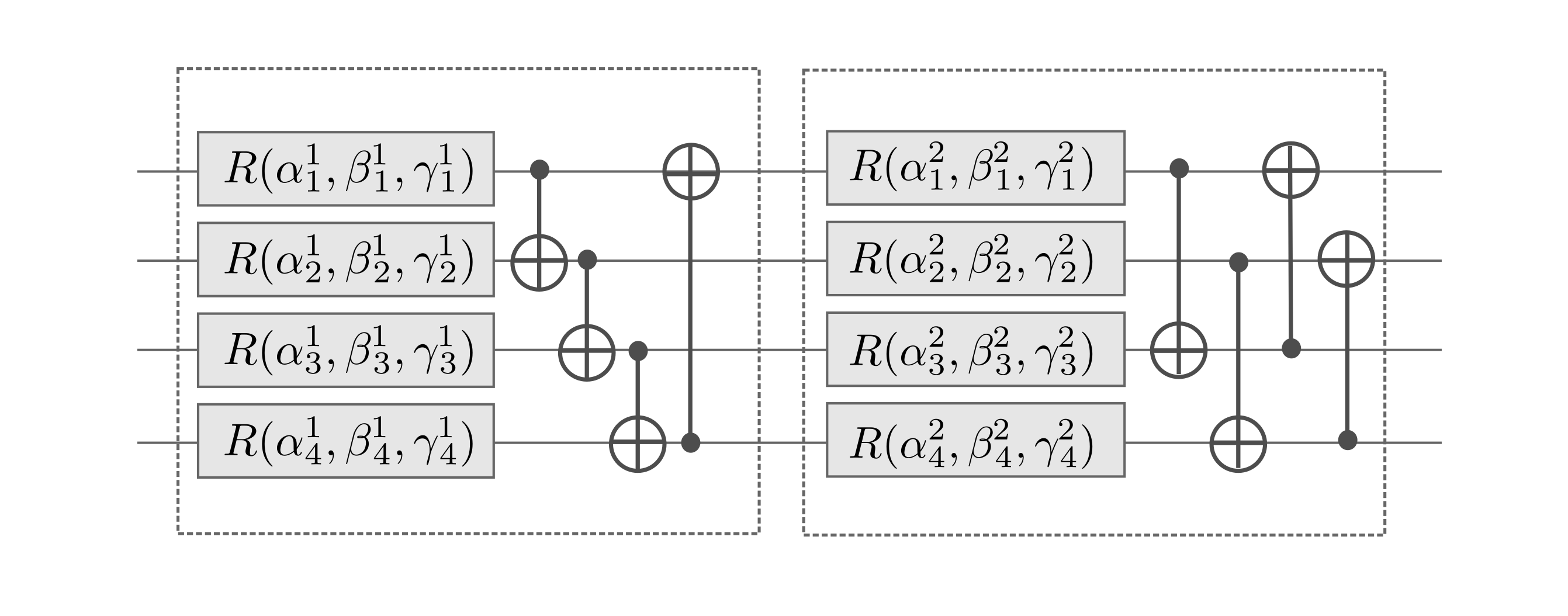}
	\caption{Strongly Entangled Circuit (\href{https://docs.pennylane.ai/en/stable/code/api/pennylane.StronglyEntanglingLayers.html}{Penny Lane})}
	\label{StronglyEnt}
\end{figure}

The architecture for the proposed HQGCNN is illustrated in Fig. \ref{QGNN_arch}, where the GCNN part is combined with a Quantum MLP layer. The drawings in the blocks are introduced for illustrative purposes only, and their description is detailed in the following. \\
After loading in input the SODA dataset, $V_i$, i = 0, N, augmentation of the dataset has been carried on as explained in the previous section and
the 
$Z_i^{0}$, i = 0, N, have been generated. After that, 
the node embedding $Z_i^{l}$ for each generic  GCN layer $l$ is generated, with l = 0, L. 
The relationship that binds 
the node embedding at one level and the next, is expressed by the equation:
\begin{equation}
    Z^{(l)} =  \sigma (A Z^{(l-1)}W^l)
\end{equation}
where $\sigma$ is the activation function, A is the adjaceny matrix, W is a learnable parameter. The \textit{edge structure} of the graph is learnable and for this reason the adjacency matrix is not fixed.
\noindent
At the exit from the last layer, the components of the vector $Z^L$ are aggregated through an aggregation function $g$. \\Finally, through the MLP layer the ONI is forecasted, with an appropriate loss function, the mean squared error (MSE) between the predicted and the Ground Truth (GT) ONI index, minimized. In the proposed architecture, the classical version of the MLP layer has been substituted with a Quantum version and different circuit combinations have been trialed. 
Namely, various gate operations can be chosen to encapsulate a quantum representation of classical data \cite{abbas2021power}.\\ In the proposed method, we employed an \href{https://docs.pennylane.ai/en/stable/code/api/pennylane.AmplitudeEmbedding.html}{Amplitude Encoding} to enhance data representation. This is a strategic choice driven by the large size of data since it encodes $2^n$ features into the amplitude vector of n qubits. For the Amplitude Encoding,  three different circuits have been tested: a) Basic Entangled Circuit, b) Strongly Entangled Circuit, and c) Random Circuit. We also tested different configurations of Layers and Qubits, mostly built on the guidelines proposed in \cite{sebastianelli2023quantum}, even if additional simulations are necessary. Based on preliminary results, it seems that the configuration that works better is with the Strongly Entangled Circuit, with 4 layers and 6 qubits, as shown in Fig. \ref{StronglyEnt}. Currently, predictions have only been made for 1 and 3 months ahead (n = 1, 3), but work is in progress to test the algorithm for various values of n. From these preliminary results, it appears that we surpass the SOTA, as shown in Table \ref{tab:res} where the values of the \textbf{all-season correlation skill}, used as an evaluation metric, are reported for some different models. It is worth underscoring that the authors have replicated the results of the Graphino model by using the code released in \cite{cachay2021world}, and these results have been reported in the table. Furthermore, we have achieved significant results by training the model for only 5 epochs (SOTA works typically use 50 epochs), so this achievement is noteworthy. In any case, future work will consider increasing the number of epochs for further evaluation.
\\
By now interested readers can refer to the \href{https://alessandrosebastianelli.github.io/hybrid_quantum_models/hqm.html}{Hybrid Quantum Models} library created by the authors, while the developed code for the proposed HQGCNN  will be made available after the paper publication.


\begin{table}[!ht]
    \centering
    \caption{The proposed HQGCNN, named "QGraphino", outperforms the SOTA 
    for a forecast of one and three months. The \textbf{all-season correlation
skill} was measured on the held-out GODAS test set (1984-2017) with the same setup as \cite{cachay2021world}.\\}\label{tab:res}
    \begin{tabular}{lcc}
    \toprule
    Model & n = 1 & n = 3 \\
    \midrule
    (a) Graphino \cite{cachay2021world} & 0.963 &  0.846 \\
    (b) SINTEX-F \cite{luo2008extended} & 0.895 & 0.840 \\
    (c) CNN \cite{ham2019deep} & 0.942 & 0.876 \\
    (d) \textbf{QGraphino} & \textbf{0.974} & \textbf{0.884}  \\
    \bottomrule
    \end{tabular}
\end{table}

\section{Discussions and Conclusions}
A Hybrid Quantum Graph Convolutional Neural Network (HQGCNN) for ONI forecasting, named by the authors as \textbf{QGraphino}, has demonstrated to outperform existing SOTA models with an \textbf{all-season correlation skill} of 0.97 in one-month prediction and 0.89 in three-month prediction. Featuring a Strongly Entangled Circuit with 4 layers and 6 qubits, our QGraphino achieves this performance with just 5 training epochs, showcasing high efficiency. The model's transformative potential in Earth and Atmospheric sciences is evident, emphasizing the promise of quantum-enhanced machine learning. The developed code will be shared post-publication.

\vspace{-0.4cm}
\balance
\bibliographystyle{IEEEtran}
\bibliography{strings,refs}

\end{document}